\documentclass[letter]{aa} 
\usepackage{natbib}
\usepackage{lscape}
\usepackage{rotating}
\usepackage{graphicx}
\usepackage{amsmath}
\usepackage{enumerate}
\usepackage{txfonts}

\newcommand{\gta}{\lower 0.5ex\hbox{$ \buildrel>\over\sim\ $}}
\newcommand{\lta}{\lower 0.5ex\hbox{$ \buildrel<\over\sim\ $}}

\newcommand{\teff}{$T_{\mathrm{eff}}$}
\newcommand{\Msun}{$M_{\mathrm{\odot}}$}

\usepackage{subcaption} 
\usepackage{url}
\begin{document}
%
\title{Asteroseismic test of rotational mixing in low-mass white dwarfs}
  
   \author{A. G. Istrate
          \inst{1}\fnmsep\thanks{e-mail: istrate@uwm.edu},
          G. Fontaine \inst{2},
          A. Gianninas \inst{3}, 
          L. Grassitelli \inst{4},
          P. Marchant\inst{4},
          T. M. Tauris\inst{5,4}
          \and
           N. Langer\inst{4} }

   \authorrunning{Istrate et. al}
   \titlerunning{Asteroseismic test of rotational mixing in low-mass white dwarfs}

   \institute{Center for Gravitation, Cosmology, and Astrophysics,
  Department of Physics, University of Wisconsin-Milwaukee, P.O. Box
  413, Milwaukee, WI 53201, USA 
   \and
   D\'epartement de Physique, Universit\'e de Montr\'eal,
  C.P. 6128, Succursale Centre-Ville, Montr\'eal, QC H3C 3J7, Canada
  \and
  Homer L. Dodge Department of Physics and Astronomy,
  University of Oklahoma, 440 W. Brooks St., Norman, OK 73019, USA
  \and
    Argelander-Institut f\"ur Astronomie, Universit\"at Bonn, Auf
              dem H\"ugel 71, 53121 Bonn, Germany
  \and
   Max-Planck-Institut f\"ur Radioastronomie, 
              Auf dem H\"ugel 69, 53121 Bonn, Germany 
              }

   \date{Received October, 2016}

 
  \abstract{ We exploit the recent discovery of pulsations in mixed-atmosphere
(He/H), extremely low-mass white dwarf precursors (ELM proto-WDs) to
test the proposition that rotational mixing is a fundamental process in
the formation and evolution of low-mass helium core white
dwarfs. Rotational mixing has been shown to be a mechanism able to
compete efficiently against gravitational settling, thus accounting
naturally for the presence of He, as well as traces of metals such as Mg
and Ca, typically found in the atmospheres of ELM proto-WDs. Here we
investigate whether rotational mixing can maintain a sufficient amount
of He in the deeper driving region of the star, such that it can fuel,
through  He{\sc{ii}}-He{\sc iii} ionization, the observed pulsations in this
type of stars. Using state-of-the-art evolutionary models computed with
MESA, we show that rotational mixing can indeed explain qualitatively
the very existence and general properties of the known pulsating,
mixed-atmosphere ELM proto-WDs. Moreover, such objects are very likely to pulsate again during their final WD cooling phase.    }

\keywords{ asteroseismology --- binaries: close --- stars: evolution ---stars: white dwarfs }   

   \maketitle

\section{Astrophysical context}

\cite{gianinas16} recently reported the discovery of pulsations in three
mixed-atmosphere, extremely low-mass white dwarf precursors (ELM
proto-WDs). Their location in the $\log g-T_{\mathrm{eff}}$\footnote{\textit{J0756+6704}: \teff = 11,640 $\pm$ 250 K, log $g$ = 4.90 
$\pm$ 0.14, $X$(He) = 0.50 $\pm$ 0.20; \textit{J1141+3850}: \teff = 11,290 $\pm$ 210 K, log $g$ = 4.94 $\pm$ 0.10, $X$(He) = 0.54 $\pm$
0.14; \textit{J1157+0546}: \teff = 11,870 $\pm$ 260 K, log $g$ = 4.81 $\pm$ 0.13, $X$(He) = 0.53 $\pm$ 0.20.} diagram and  the detected periods are similar to those of the first  discovered pulsating ELM proto-WDs WASP
0247-25B \citep[\teff = 10,840 $\pm$ 300 K, log $g$ = 4.576 $\pm$ 0.011;][]{maxted2013}  and
WASP 1628+10B \citep[\teff = 9,200 $\pm$ 600 K, log $g$ = 4.49 $\pm$ 0.05;][]{maxted2014}. 
 It is expected that the type of the  pulsation driving is the same in both types of systems  \citep[see, e.g.,][]{jeffery2013, corsico2016}. While the (likely) presence of He in the atmosphere of the two WASP systems has yet to be confirmed\footnote{a difficult task as the  light of the A-type companion dominates the optical spectrum}, the results of \cite{gianinas16} represent the first empirical evidence that pulsations
in relatively hot ELM proto-WDs can only occur when  a significant amount of He is present in their atmospheres. We disregard  here the two cool-ELM proto-WD  candidates proposed by \cite{corti2016}, and also  the system discussed by  
 \cite{zhang2016}, as their nature is  currently unclear. 

    Helium is the ingredient needed to drive pulsations in a regime of effective temperature well above the
blue edge of the ZZ Ceti instability strip, as well as its extension into the low-gravity domain  \citep[e.g.,][]{steinfadt2010,corsico2012, van_grootel2013}. The  ZZ Ceti instability strip  only contains pure H atmosphere (DA) WDs for which pulsation driving is confined to the regions of partial ionization of H. \cite{vfb2015}  showed that a full continuum of instability strips, from the cooler pure H ZZ Ceti to the hotter pure He V777 Her domain, is obtained uniquely as a function of the He/H envelope ratio along the WD
cooling tracks. In this case, pulsation driving is due to the combined effects of partial ionization of H and He.
Non-adiabatic stability analysis of simple envelope
models  indicates that the pulsations in the newly discovered three mixed-atmosphere proto-WDs are caused
mostly by a standard $\kappa$-mechanism associated with the
second ionization of He, in conjunction with some convective driving \citep{gianinas16}.

Regarding the evolution of ELM WDs, \cite{istrate2016} investigated the combined effects of
rotational mixing and diffusion processes\footnote{gravitational settling, thermal diffusion,  and chemical diffusion.} on the (proto-) WDs that are formed through the low-mass X-ray binary channel. After the end of the mass-transfer phase, the envelope of the newly formed proto-WD contracts significantly, rotating thus faster than  the  helium core. This gives rise to rotational instabilities, with Eddington-Sweet circulation  being the main process responsible for the mixing of material.\\
 In particular, rotational mixing was shown to be a mechanism able to compete efficiently against gravitational settling, which would otherwise lead to the formation of a pure H atmosphere on a very short timescale \citep[e.g.,][]{althaus2001, althaus2013}. This is  in agreement with what has been observed in ELM proto-WDs, where substantial amounts of He, as well as traces of metals such as Mg and Ca, are often detected  in their atmospheres
\citep{kaplan2013,gianinas2014,hermes2014,latour2016}. While the influence of radiative levitation  has yet to be investigated quantitatively in these stars, it is certain that it cannot explain the detected relatively large amounts
of He.  This is because radiative levitation is only able to support traces of various elements in a dominant
background  because of to line saturation. Considering this, rotational mixing is most  likely a fundamental process that should be taken into account in the evolution  of ELM (proto-) WDs  with important consequences in the  future asteroseismological studies of this class of objects.

In this Letter we demonstrate that the new available evolutionary models can  account for the existence of the observed pulsating mixed-atmosphere ELM proto-WDs. 

\section{Results}
We identified several evolutionary sequences from \cite{istrate2016} and recomputed them  to achieve a higher density of models in the region of interest. These new tracks have been calculated using the publicly available binary stellar evolution code MESA, version 7624 \citep{mesa1, mesa2, mesa3}. The details regarding
the input physics are the same as in \cite{istrate2016}. It should be noted here that there are still many uncertainties in our theoretical understanding of rotational mixing. For instance, the effects of the Eddington-Sweet circulation \citep{heger2000} and of the Tayler-Spruit dynamo \citep{spruit2002} are  both not well constrained. Models including magnetic fields  reproduce the angular momentum content of stellar cores  much better \citep{suijs2008}, although the validity of this process has been questioned \citep{zahn2007}. Additional effects such as gravity waves \citep[which were not considered in the models of ][]{istrate2016} might also play an important role in the transport of angular momentum \citep{fuller2014}.

We carried out a stability analysis of these evolutionary models using
 the Montr\'eal pulsation codes
\citep{brassard1992,fontaine1994}. For each retained model, we
investigated the stability of all pulsation modes in the range of
periods from 50 to 4000 s and with degree index $\ell$ = 0, 1, and 2.
When present, convection was handled through the so-called frozen
flux approximation, for which the perturbations of the convective flux
are ignored. To the extent that convective driving does not dominate the
excitation process in ELM proto-WD models, this was deemed sufficient
\citep{van_grootel2012, van_grootel2013, vfb2015}; see also Sect.~\ref{discussion} below. \

\begin{figure}
 \includegraphics[width=\linewidth]{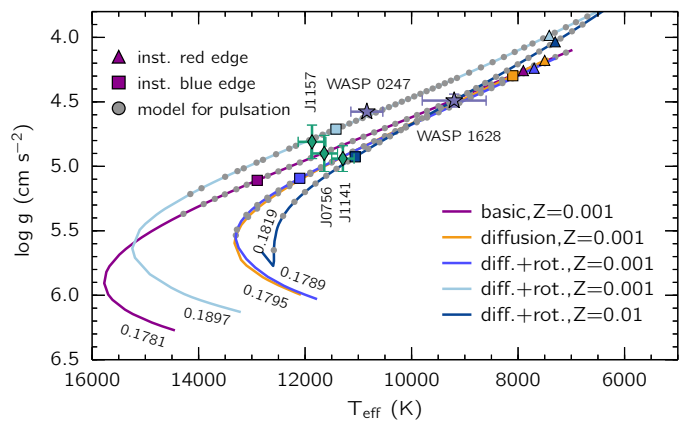}
 \caption{ELM WD evolutionary tracks representative of the observed proto-WD pulsators used for pulsation analysis. The mass (M$_{\odot}$)  is indicated by the numbers next to each track.}\label{fig:fig1b} 
\label{fig:loggteff} 
\vspace{-8px} 
 \end{figure}

\begin{figure}
 \includegraphics[width=\linewidth]{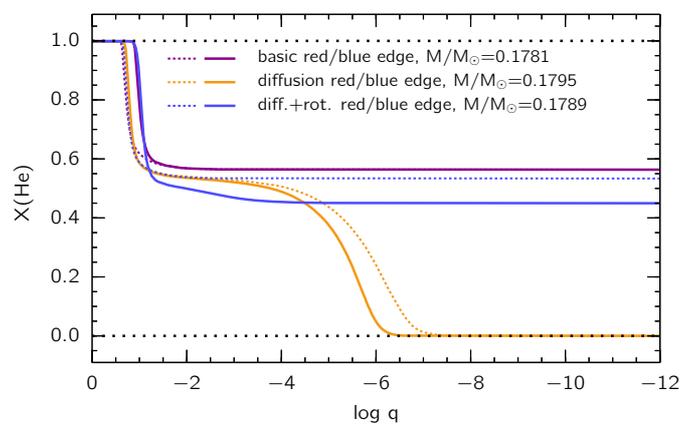}
 \caption{Helium mass fraction, $X$(He), as a function of the fractional
   mass depth, log $q$ $\equiv$ log ($1-M(r)/M_*$) for the red and blue edge profiles for the models in \textit{basic}, \textit{diffusion}, and \textit{rotation+diffusion} configurations. The assumed metallicity is Z=0.001.} 
\label{fig:helium} 
\vspace{-8px} 
 \end{figure}
 
  Figure~\ref{fig:loggteff} shows the five evolutionary tracks used in this work in the $\log g-T_{\mathrm{eff}}$ diagram.
Following \cite{istrate2016}, we investigated the pulsational behavior of three different configurations, with roughly the same WD mass, computed  for a metallicity of $Z=0.001$: (i)  \textit{basic}, where no element diffusion nor rotational mixing are included, (ii) \textit{diffusion}, 
in which element diffusion operates, and (iii) \textit{diffusion+rotation}, in which both element diffusion and rotational mixing are present. The chemical stratification of the envelope  in the case of the \textit{basic} configuration is set by the outcome of the binary evolution and it is changed only by the hydrogen shell burning.   
The choice of metallicity is partly motivated by the fact that both  J0756+6704 and J1141+3850 are probably halo stars \citep{gianninas2015,brown2016_merger}\footnote{In the case of J1157+0546, the population membership is still unknown.}. Additionally, we  also study  the influence  of the  WD mass and metallicity on the excited modes.

Figure~\ref{fig:helium} illustrates that rotational mixing plays a key role in maintaining relatively large
amounts of He in the envelope of the models on the proto-WD branch. In contrast, pure diffusion
leads very early on to the formation of a pure H envelope. From a pulsation point of view, this is a fundamental difference as He is
needed, through its second ionization stage, for exciting pulsation modes via a $\kappa$-mechanism in the domain where the pulsating ELM proto-WDs are found, that is, in a regime of temperature where H is
completely ionized and cannot help in the destabilization process.  
\begin{figure*}
\centering  
\vspace{-12px}
  \begin{subfigure}{.45\textwidth}
  \includegraphics[width=\linewidth]{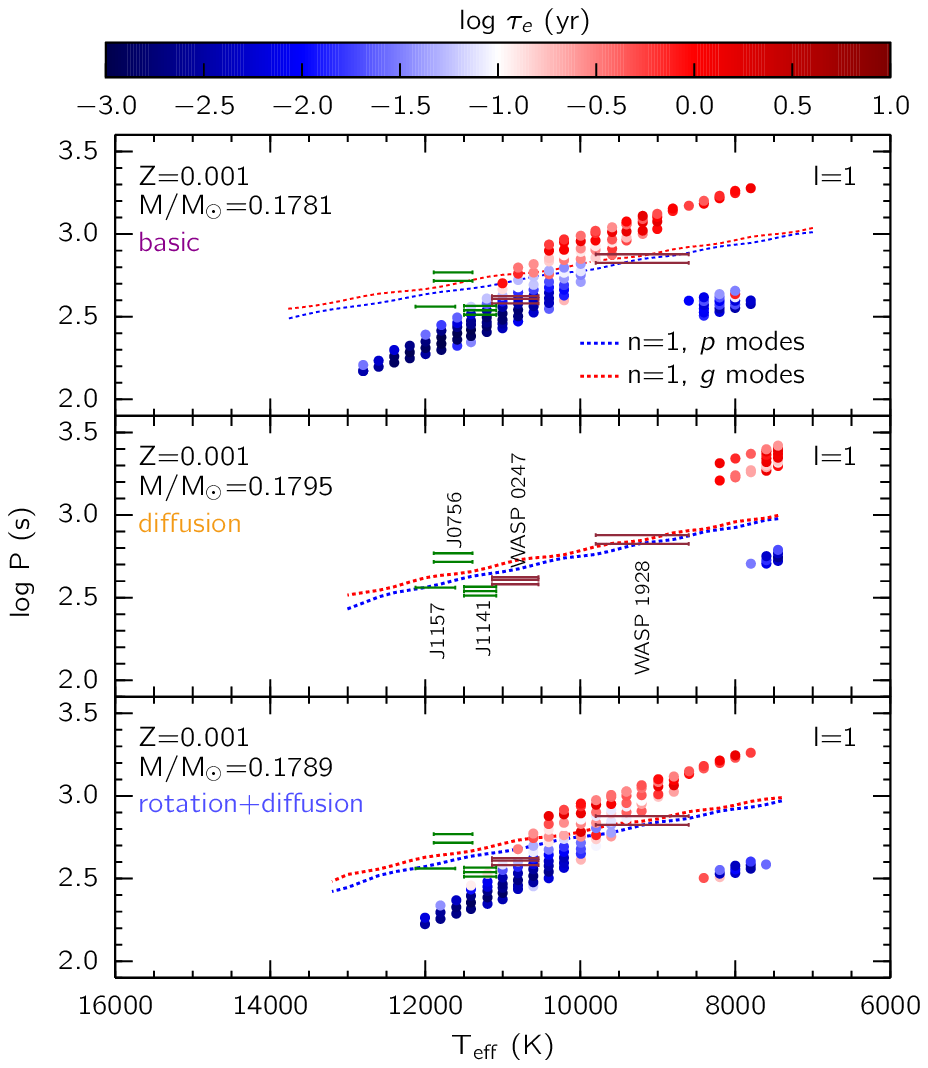}  
  \caption{}  
  \end{subfigure}
  \begin{subfigure}{.45\textwidth}
  \includegraphics[width=\linewidth]{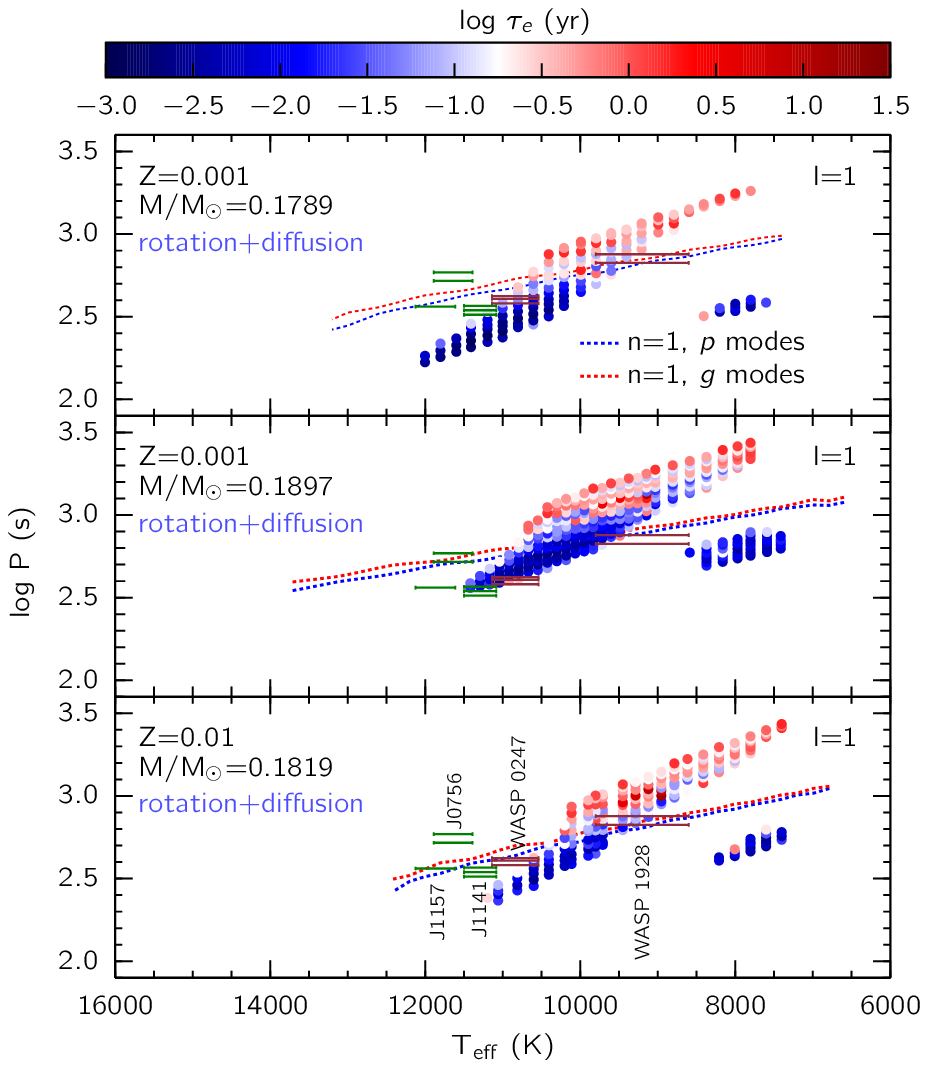}
    \caption{} 
  \end{subfigure}
\vspace{-12px}
\caption{ Calculated period spectrum for the excited modes for each of the retained models represented by the gray circles in Fig.~\ref{fig:loggteff}. The color code indicates the logarithm of the e-folding time, $\tau_{e}$. The blue dotted curve defines the upper limit (radial order $n=1$) of the $p$-mode spectrum, while the red dotted curve corresponds to the lower limit (radial order $n=1$) of the $g$-mode spectrum. The green (magenta) lines denote the three observed mixed-atmosphere pulsators (two WASP pulsators).  Left panel: comparison between the \textit{basic}, \textit{diffusion}, and \textit{diffusion+rotation} for roughly the same WD mass; right panel:  comparison of different WD masses and metallicities in the \textit{rotation+diffusion} configuration.}
  \label{fig:modes}  
\end{figure*}
We note  that the region of damping or driving is typically located in the
range $-$7 $\gta$ log $g$ $\gta$ $-$11 in the envelope of ELM  proto-WD
models \citep{corsico2016}. The \textit{diffusion+rotation} (\textit{basic}) model
 predicts that the He content
in the damping or driving region is the same as in the outermost layers (see Fig.~\ref{fig:helium}). In this case, the
atmospheric He abundance, as determined through standard spectroscopic
techniques, provides a direct measure of the quantity of that element in
the much deeper, unobservable damping or driving region. This can become a
very useful diagnostic tool. 

Figure~\ref{fig:modes} summarizes our stability analysis results for $\ell=1$  modes\footnote{The excited $\ell$
= 1 modes map essentially the same domain in the period-effective temperature plane as the $\ell$ = 0 and $\ell$ = 2 modes.}.  For the \textit{basic} model, the He content in the envelope remains
fixed at a value of $X$(He) $\simeq$  0.57, which leads to a maximum extension of the instability domain. In the early cooler phase (\teff $<$ 9,000 K) of ELM proto-WD  evolution, two instability islands can clearly be seen, one corresponding to $p$-modes (acoustic modes), and the
other to $g$-modes (gravity modes). With increasing T$_{\mathrm{eff}}$, a single broad band
of excited periods persists, initially consisting of $g$-modes, and
finally consisting of $p$-modes at the blue edge (\teff $\simeq$ 12,800 K). 

In the \textit{diffusion} model, pulsational instabilities are found only in the cooler models (\teff $<$ 8,200 K), which are already characterized by a pure H composition in the driving region. Although in reality unimpeded diffusion is  unlikely to occur, such configurations would correspond to cool pulsating ELM proto-WDs lying in the extension of the classical ZZ Ceti instability strip. This case of pure diffusion, however, is in direct conflict with the existence of
the known pulsating ELM  proto-WD stars, which are found at much higher effective temperatures. This has been pointed out previously by \cite{corsico2016}. 

In the case of the \textit{diffusion+rotation} model, we find that rotational mixing is able to maintain, against gravitational settling, a sufficient amount of He in the envelope of ELM proto-WDs to drive pulsations in the regime of effective temperatures in which the observed pulsators are found. Specifically, the initial helium content in the envelope is $X$(He) $\simeq$ 0.54 and only slowly drops to $X$(He) $\simeq$ 0.46 (see Fig. \ref{fig:helium}) by the time the model has
reached the blue edge of the instability region. The latter is located at \teff $\simeq$ 12,000 K, cooler than that of the \textit{basic} sequence as expected, but still providing a very adequate coverage of the
observed instability domain. 

 The predicted domain of instability in the period-effective temperature  plane is very sensitive to the stellar mass,  but also depends on the  assumed metallicity. The dependence of the computed domain of instability on stellar mass has been described in detail by \cite{corsico2016}. The main effect of increasing the mass, all other  things being equal, is to displace the instability domain to a  somewhat cooler and longer-period region. For instance, the blue edge is shifted from \teff~ $\simeq$ 12,000 K (0.1789~\Msun) to \teff~ $\simeq$ 11,400 K (0.1897~\Msun), while the red edge is lowered by
 $\simeq$ 200 K. This is  directly related to the lower initial amount  of He left in the envelope of the proto-WD after the end of the mass-transfer phase.  With an increased metallicity, $Z$ = 0.01, we find that the blue edge  is instead lowered to \teff~ $\simeq$ 11,100 K. This behavior is again caused by the lesser initial amount of He that is available  at the outset of the final evolutionary phase. For a given mass, binary evolution leads to an  envelope composition less enriched in He if the assumed metallicity is higher \citep{istrate2016}.  \\
  Summarizing, we find that evolutionary models that include rotational mixing provide a natural explanation for the observed pulsational instabilities in the ELM proto-WD regime. Additionally, we obtain that the qualitative match between the predicted domain of instability and the observations is better for the sequence with the relatively low mass of 0.1789 \Msun~ and  metallicity of $Z= 0.001$.  

\section{Discussion}\label{discussion}
 To compare the detected pulsation periods with the computed  periods, we need to fold in the important sensitivity of the pulsation  results on the WD mass. Figure~\ref{fig:loggteff} shows that both the  0.1789 \Msun~ and the 0.1819
\Msun~ sequence  {\sl might} serve as the basis for a seismic model for both WASP 1628+10B and J1141+3850.
Likewise, the 0.1897 \Msun~sequence {\sl might} be of interest for  WASP 0247$-$25B and J1157+0546. In this connection, the top panel in Fig.~\ref{fig:modes}b indicates that the two periods detected in WASP~1628+10B \citep{maxted2014} and the three pulsations detected in J1141+3850 \citep{gianinas16} correspond rather well to predicted
values, taking into account the uncertainties on the estimates of the effective temperature of these stars. In addition, for
J1141+3850, we have an estimate of the spectroscopic abundance of helium, $X$(He) =  $0.54~\pm$ 0.14 \citep{gianinas16}. While the
uncertainties are relatively large, this agrees with the range of helium abundance expected in the envelope of the 0.1789
\Msun~ models. 

The agreement between the two observed periods in WASP~1628+10B and the predicted values improves slightly 
by considering the $Z$ = 0.01, 0.1819 \Msun~ sequence. In contrast, the predicted excited periods are somewhat shorter than those observed in J1141+3850. We also find that the results of the  0.1897 \Msun~ sequence  are in
qualitative agreement with the three periods detected in WASP~0247$-$25B, while the predicted blue edge of the instability domain
falls slightly short of the location of J1157+0546. The single detected period of 364 s is, however, consistent with the calculated periods at that blue edge. The atmospheric value of the He abundance in J1157+0546 is $X$(He) = $0.53~\pm$ 0.20 \citep{gianinas16}; this is not well constrained, but nevertheless consistent with the expected values shown in  Fig.~\ref{fig:helium}. 

With perhaps one exception, we thus find  an overall good agreement between the predictions of the rotation+diffusion
approach and the general properties of the known pulsating ELM proto-WD stars. The exception is J0756+6704, which appears to be somewhat problematic as a result of the combination of a relatively high estimated effective temperature
with two long detected periods at 521 and 587 s \citep{gianinas16}.  
  
Taking into account the current revision of the convective efficiency for the atmospheric modeling of DA WDs from ML2/$\alpha$=0.8 to ML2/$\alpha$=0.7 (Pierre Bergeron 2016, private communication), we computed a new grid of models for the atmospheric modeling and refitted the spectra for the three pulsators. As a result of less efficient convective transport, we expect that the estimated effective temperatures to be revised downward. However, there is practically  no convective flux in the atmospheric layers
of these relatively hot proto-WDs, and therefore the estimated effective temperatures are not affected by the change in convective efficiency.  Nevertheless, there is still a non-negligible convective flux in the much deeper layers, near the driving
region in evolutionary models of such stars. 

 There are several possibilities that could help solve the apparent conflict between observations and theory in the case of J0756+6704. First, the efficiency of rotational mixing is poorly constrained; a higher efficiency would extend the  region of instability to higher T$_{\mathrm{eff}}$. However, the effect is limited, and can be measured by using configurations in the \textit{basic} mode. Second, the band of excited periods can potentially be widened and the blue edge can be pushed to higher temperatures when perturbations of the convective flux are taken into account. This has been demonstrated for ZZ Ceti stars for which pulsational instabilities are due to pure convective driving \citep{van_grootel2013}. As indicated above, most of the driving in pulsating ELM proto-WD  models is due to a $\kappa$-mechanism associated with  He{\sc{ii}}-He{\sc iii} ionization, but convective driving also contributes. It remains to be seen, with detailed calculations, whether these effects would be important in a ELM proto-WD context. And third, the convective efficiency assumed in our current evolutionary models has an additional, if indirect, effect on the pulsation properties: it
affects the stratification of the envelope. In the present case, the efficiency assumed in the construction of the models
\citep[see][]{istrate2016} is lower than the ML2/$\alpha$=1.0 version that has been calibrated in the work of \cite{van_grootel2013}  for ZZ Ceti stars. Using the latter version would increase the contribution of convective driving and, presumably, widen the band of excited periods and the width of the instability strip. Moreover, the additional pressure arising from the turbulent convective motion can be as high as a few percent, which could lead to more structural differences and observational effects \citep{grassitelli2015}.
 We are thus optimistic that the
case of J0756+6704 can be solved. 

We conclude that our current evolutionary models of ELM WDs appear indeed quite compatible with the very existence of pulsating ELM proto-WD stars. Rotational mixing is able to oppose gravitational settling and maintain a sufficient amount of He in the envelopes of such stars for the models to develop pulsational instabilities with characteristic periods  that generally agree well with the detected periods.  By the time such an object enters its cooling track and  crosses the blue edge of the ZZ Ceti instability strip, gravitational settling dominates  rotational mixing. Thus, the star has developed a pure H 
envelope  that is able to drive pulsations again, but, this time, through pure H ionization, and as a cool, low-mass ZZ Ceti (ELMV) star
\citep{hermes2013}. Hence, a pulsating ELM  proto-WD star is very likely to pulsate again during its final WD cooling phase.

\begin{acknowledgements}
A.G.I acknowledges support from the NASA ATP program through NASA grant
NNX13AH43G. This work was also supported in part by the NSERC Canada
through a research grant awarded to G.F. The latter also acknowledges the 
contribution of the Canada Research Chair Program. A.G. gratefully
acknowledges the support of the NSF under grant AST-1312678, and NASA
under grant NNX14AF65G.
\end{acknowledgements}

\bibliographystyle{aa}
\bibliography{ifggptl}


\end{document}